\documentclass{pasj00}
\draft
\usepackage{longtable}

\newcommand{\cjaa}{Chinese J. Astron. Astrophys.}
\newcommand{\actaa}{AcA }
\newcommand{\nar}{NewA. Rev.}
\newcommand{\na}{NewA.  }

\begin{document}
\SetRunningHead{Author(s) in page-head}{Running Head}

\title{Photometric Investigation of the Contact Binary GU Ori with High Metallicity}



%
 \author{Zhou \textsc{X.}\altaffilmark{1,2,3,5},
         Qian  \textsc{S.-B.}\altaffilmark{1,2,3,4},
         Soonthornthum \textsc{B.}\altaffilmark{5},
         Poshyachinda \textsc{S.}\altaffilmark{5},
         Zhu \textsc{L.-Y.}\altaffilmark{1,2,3,4},
         Liu \textsc{N.-P.}\altaffilmark{1,2,3},
         Sarotsakulchai \textsc{T.}\altaffilmark{1,2,3,4},
         Fang \textsc{X.-H.}\altaffilmark{1,2,3,4}}
\altaffiltext{1}{Yunnan Observatories, Chinese Academy of Sciences (CAS), P.O. Box 110, 650216 Kunming, P. R. China}
 \email{zhouxiaophy@ynao.ac.cn}
\altaffiltext{2}{Key Laboratory of the Structure and Evolution of Celestial Objects, Chinese Academy of Sciences, P. O. Box 110, 650216 Kunming, P. R. China}
\altaffiltext{3}{Center for Astronomical Mega-Science, Chinese Academy of Sciences, 20A Datun Road, Chaoyang Dis-trict, Beijing, 100012, P. R. China}
\altaffiltext{4}{University of the Chinese Academy of Sciences, Yuquan Road 19\#, Sijingshang Block, 100049 Beijing, P. R. China}
\altaffiltext{5}{National Astronomical Research Institute of Thailand, 260  Moo 4, T. Donkaew,  A. Maerim, Chiangmai, 50180, Thailand}
\KeyWords{binaries:photometric -- stars: formation -- stars: individual: GU Ori} 

\maketitle

\begin{abstract}
GU Ori was observed with the 1m telescope at Yunnan Observatories in 2005. To determine its physical properties, the Wilson-Devinney program is used. The results reveal that GU Ori is a W-subtype shallow contact binary with a more massive but cooler star 2. The mass of its two component stars are estimated to be $M_1 = 0.45M_\odot$, $M_2 = 1.05M_\odot$. The O'Connell effect was reported to be negative on the light curves observed in 2005. However, it changed to a positive one on the light curves observed from 2011 to 2012. The mean surface temperatures of star 2 ($T_2$) determined by the two sets of light curves were different, which may result from stellar activities. The O - C diagram shows that the period of GU Ori is decreasing at a rate of $dP/dt=-6.24\times{10^{-8}}day\cdot year^{-1}$, which may be caused by mass transfer from star 2 to star 1 with a rate of $\frac{dM_{2}}{dt}= - 2.98\times{10^{-8}}M_\odot/year$. GU Ori is a contact binary with quite high metallicity.
\end{abstract}

\section{Introduction}

In the Universe, over $60\,\%$ of stars are in binary or multiple systems. However, only a very small part (nearly $0.2\,\%$) of them that eclipses can be observed from our Earth \citep{2004NewAR..48..647G}. Eclipsing binaries are ideal targets to determine absolute parameters of stars when light curves and radial velocity curves are combined. The evolutionary of close binaries is more complex than those of single stars, since the evolution of components will be highly affected by each other. \citet{2018ApJS..235....5Q} divide the W UMa-type contact binaries into three groups: (1) long-period massive EWs with the same metallicities with lower Log g than those of EAs with the same orbital period; (2) short-period less-massive EWs that have lower metallicities and higher Log g; (3) EWs have higher metallicities that may be contaminated by the material from the evolution of compact companion objects. In the present work, we focus on the contact binary GU Ori with high metallicity ($[Fe/H] = 0.31$). It belongs to extreme Populaton I stars and locates in the spiral arm of the Milky Way galaxy. \citet{2018ApJS..235....5Q} assume that they may be contaminated by the material from compact objects nearby. Many contact binaries are found to be accompanied with tertiary components \citep{2015AJ....150...83Z,2016AdAst2016E...7Z,2016ApJ...817..133Z}. \citet{2018ApJS..235....6T} updates the catalog of multiple stars. It is supposed that the tertiary component may accelerate the orbital evolution of the host binary through the Kozai mechanism \citep{1962AJ.....67..591K,2007ApJ...669.1298F} and shorten the pre-contact time scale of close binaries.

GU Ori was listed in the General Catalog of Variable Stars (GCVS) \citep{2017ARep...61...80S} as an Algol type variable star. \citet{1985JAVSO..14...12S} determined its period to be 0.470681 days and concluded that the period of GU Ori isn't constant. However, it is listed as a W UMa type contact binary in the International Variable Star Index (VSX) \citep{2006SASS...25...47W}, with photometric magnitude range from 12.5 to 13.3 mag in $V$ band, and period of 0.4706769 days. \citet{2017PASJ...69...69Y} claimed its spectral type to be G0V and published $V$ and $R_c$ light curves' solution. The spectrum of GU Ori was obtained on January 3rd, 2014 by the Large Sky Area Multi-Object Fiber Spectroscopic Telescope (LAMOST) \citep{2012MNRAS.419.3406S,2012RAA....12.1197C,2012RAA....12..723Z,2012RAA....12.1243L,2015RAA....15.1095L,2012RAA....12..735D}, which determined its spectral type to be G1, effective temperature to be 6050 $K$, and metallicity ([Fe/H]) to be 0.31. In the present work, the light curves and period variations of GU Ori are investigated to understand the theory of formation on contact binaries with high metallicity.

\section{Photometric Observations}

\begin{figure}[!ht]
\begin{center}
\includegraphics[width=8cm]{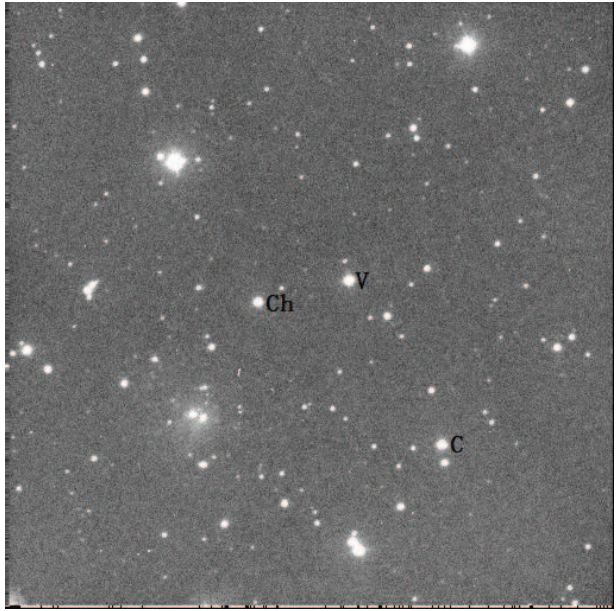}
\caption{The observational image of GU Ori on Feb. 6th, 2005.}\label{image}
\end{center}
\end{figure}

GU Ori was observed on February 6th and 9th, 2005 with the 1.0 m reflecting telescope at Yunnan Observatories, Chinese Academy of Sciences, and the $B$, $V$ and $R_c$ filters (Johnson-Cousins filter system) were used. The 1m telescope has a field of view about 7.3 square arc-minutes. The exposure time was 140s for all images. A total of 79 images in $B$ band, 78 images in $V$ band, 77 images in $R_c$ band were obtained. One of the observational images is displayed in Fig. \ref{image}.

All of the observational images are reduced with IRAF. UCAC4 514-020704 and UCAC4 515-021077 are chosen as the Comparison star (C) and the Check star (Ch). Differential aperture photometry method is applied to calculate light variations of GU Ori. GU Ori (V), UCAC4 514-020704 (C) and UCAC4 515-021077 (Ch) are marked in Fig. \ref{image}. Their coordinates and V band magnitudes are listed in Table \ref{Coordinates}.

\begin{table}[!h]\small
\begin{center}
\caption{Coordinates and the $V$ band magnitudes.}\label{Coordinates}
\begin{tabular}{cccc}\hline\hline
Target             & $\alpha_{2000}$           &  $\delta_{2000}$          &  $V_{mag}$   \\ \hline\hline
GU Ori             &$06^{h}10^{m}04^{s}.621$   & $+12^\circ49'46''.63$     &  $12.657$     \\
UCAC4 514-020704   &$06^{h}10^{m}02^{s}.468$   & $+12^\circ47'52''.02$     &  $12.953$     \\
UCAC4 515-021077   &$06^{h}10^{m}08^{s}.639$   & $+12^\circ49'47''.32$     &  $14.283$     \\
\hline\hline
\end{tabular}
\end{center}
\end{table}

Two times of minimum light are determined and the averaged values are listed in Table \ref{New_minimum}. The first two columns are the observational times of minimum light and their errors. The lower-case p and s refer to the primary minimum and the secondary one. The filters used during the observations are listed in the fourth column.

\begin{table}[!ht]\small
\begin{center}
\caption{New CCD times of minimum light}\label{New_minimum}
\begin{tabular}{cccccc}\hline\hline
    JD (Hel.)     &  Error (days)  &  p/s &   Filter    &Telescope\\\hline\hline
  2453408.1405    & $\pm0.0003$    &   s  &   $BVR_C$     &    1m   \\
  2453411.2031    & $\pm0.0003$    &   p  &   $BVR_C$     &    1m   \\
\hline\hline
\end{tabular}
\end{center}
\textbf
{\footnotesize Notes.} \footnotesize 1m denotes to the 1m reflecting telescope in Yunnan Observatories, Chinese Academy of sciences.
\end{table}

\begin{figure}[!h]
\begin{center}
\includegraphics[width=12cm]{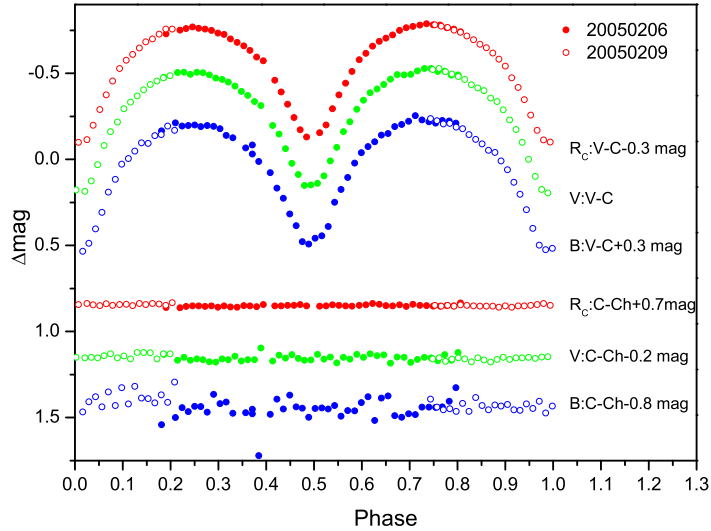}
\caption{Solid and open circles refer to data observed on Feb. 6th and 9th, 2005, respectively.}\label{LC}
\end{center}
\end{figure}

Based on the primary minimum listed in Table \ref{New_minimum} and the period given out by VSX, the phases are calculated with the ephemeris:
\begin{equation}
Min.I(HJD)=2453411.2031(3)+0^{d}.4706769\times{E}.\label{Epoch_LC}
\end{equation}
The light variations are displayed in Fig. \ref{LC}. The blue, green and red colors refer to light curves in $B$, $V$ and $R_c$ bands.
V - C means the magnitude difference between the Variable star and the Comparison star. The light curve of V - C is shifted by 0.3 mag downward in $B$ band and 0.3 mag upward in $R_c$ band. C - Ch means the magnitude difference between the Comparison star and the Check star. The light curves of C - Ch  are shifted by 0.7 mag, -0.2 mag and -0.8 mag in $B$, $V$ and $R_c$ bands, respectively. As displayed in Fig. \ref{LC}, the light curve of C - Ch in $B$ band has much larger scatter than those in $V$ and $R_c$ bands. The Check star UCAC4 515-021077 (Ch) is much fainter than GU Ori (V) and the Comparison star UCAC4 514-020704 (C). The radiant flux in $V$ and $R_c$ bands is much stronger than that in $B$ band for late-type star.

\section{Orbital Period Investigation}

Mass transfer and tertiary component are quite common in contact binary systems, which cause the periods of contact binaries to change. The O - C method is usually used to detect the period variations. The photometric observation on GU Ori has a long history and many times of minimum light have been published. They are listed in Table \ref{Minimum}.

Column 1 - heliocentric julian date of observed minimum (HJD - 2400000);

Column 2 - identification of primary (p) or secondary (s) minimum;

Column 3 - method used to determined the time of light minimum. pg means observed by photograph, and CCD refers to Charge Coupled Device;

Column 4 - cycle numbers from the initial epoch;

Column 5 - the $(O - C)$ values calculated from Equation \ref{Epoch_O-C};

Column 6 - observational error;

Column 7 - reference for the time of light minimum.

The O - C gateway\footnote{http://astro.sci.muni.cz/variables/ocgate/} is a database which collects times of minima of eclipsing binaries. GU Ori is listed in the database. The initial epoch and period published in the O - C gateway are used,
\begin{equation}
Min.I(HJD) = 2429306.135+0^{d}.470682\times{E}.\label{Epoch_O-C}
\end{equation}

\begin{figure}[!h]
\begin{center}
\includegraphics[width=12cm]{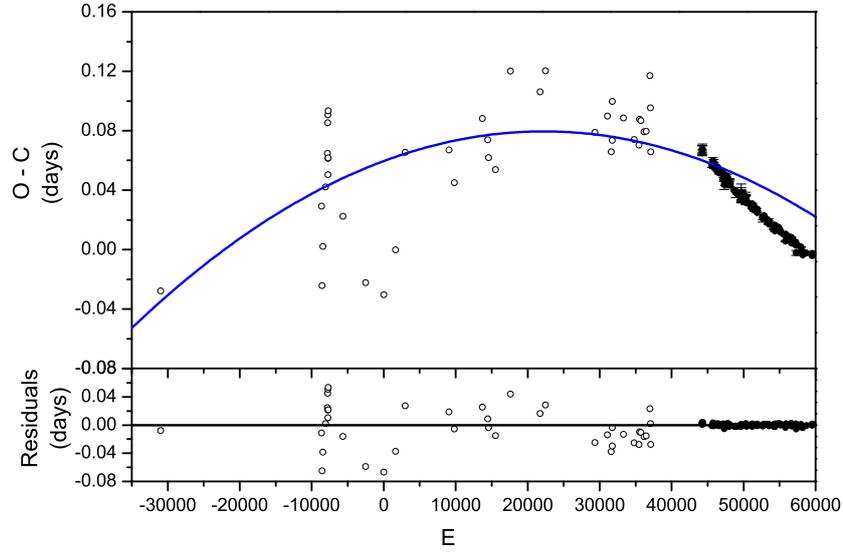}\label{O-C}
\end{center}
\caption{The open circles and solid circles represent times of light minimum observed by pg and CCD, respectively. In the upper panel, the blue line refers to the parabolic variation. The fitting residuals are displayed at the bottom.}
\end{figure}

As shown in Fig. \ref{O-C}, the period of GU Ori isn't constant. According to the result calculated by \citet{2017PASJ...69...69Y}, the parabola goes upward when only CCD data are used. However, the parabola goes downward while pg data are considered. The only reasonable explanation is that a cyclic variation exists. Therefore, long term increase (or decrease) and periodic variations are superposed on the linear ephemeris,
\begin{equation}
O - C = \Delta T_0 + \Delta P_0\times E +\frac{1}{2} \frac{dP}{dE}\times E^2 + \tau
\end{equation}
where $\Delta T_0$ and $\Delta P_0$ are corrections to the initial epoch and period, the quadratic term refers to secular variation, and $\tau$ is the periodic variation. The periodic term is explained in detail by \citet{1952ApJ...116..211I}. We assume a circular orbit in the present work.

Based on the least-squares method, the new ephemeris is determined
\begin{equation}
\begin{array}{lll}
Min. I = 2429306.19455(\pm0.00006)+0.470683791(\pm0.000000003)\times{E}
         \\-4.0226(\pm0.0069)\times{10^{-11}}\times{E^{2}}
         \\+0.02744(\pm0.00004)\sin[0.^{\circ}00662(\pm0.^{\circ}00005)\times{E}+238.^{\circ}25(\pm0.^{\circ}08)] \label{Epoch_new}
\end{array}
\end{equation}

According to the new ephemeris, the period of GU Ori is decreasing continuously at a rate of
$dP/dt=-6.24\times{10^{-8}}day\cdot year^{-1}$, and a cyclic change is revealed.

\begin{footnotesize}
\centerline{Table 3. Available times of light minima for GU Ori}\label{Minimum}
\begin{center}
\begin{longtable}{lcclrrc}
\hline\hline
JD (Hel.)    &   p/s  &   Method   &       E       &      $(O-C)$       &   Error      &  Reference        \\
(2400000+)   &        &            &               &                    &              &                   \\
\hline
\endhead
\\\hline\hline
\endfoot
\endlastfoot
\hline
14718.2600   &    p   &   	pg     &    -30993     &      -0.0278       &              &   (1)                \\
25235.4710   &    s   &   	pg     &    -8648.5    &      0.0293        &              &   (1)               \\
25272.3660   &    p   &   	pg     &    -8570      &      -0.0243       &              &   (1)                \\
25329.3450   &    p   &   	pg     &    -8449      &      0.0022        &              &   (1)                \\
25502.5960   &    p   &   	pg     &    -8081      &      0.0422        &              &   (1)                \\
25622.4070   &    s   &   	pg     &    -7826.5    &      0.0647        &              &   (1)                \\
25643.3730   &    p   &   	pg     &    -7782      &      0.0853        &              &   (1)                \\
25649.4570   &    p   &   	pg     &    -7769      &      0.0505        &              &   (1)                \\
25651.3800   &    p   &   	pg     &    -7765      &      0.0907        &              &   (1)                \\
25652.2920   &    p   &   	pg     &    -7763      &      0.0614        &              &   (1)                \\
25671.3550   &    s   &   	pg     &    -7722.5    &      0.0618        &              &   (1)                \\
25672.3280   &    s   &   	pg     &    -7720.5    &      0.0934        &              &   (1)               \\
26633.6250   &    p   &   	pg     &    -5678      &      0.0224        &              &   (1)                \\
28127.5250   &    p   &   	pg     &    -2504      &      -0.0223       &              &   (1)                \\
29306.3400   &    s   &   	pg     &    0.5        &      -0.0303       &              &   (1)                \\
30079.2300   &    s   &   	pg     &    1642.5     &      -0.0002       &              &   (1)                \\
30705.5380   &    p   &   	pg     &    2973       &      0.0654        &              &   (1)                \\
33570.5810   &    p   &   	pg     &    9060       &      0.0671        &              &   (1)                \\
33922.6290   &    p   &   	pg     &    9808       &      0.0449        &              &   (1)                \\
35757.6260   &    s   &   	pg     &    13706.5    &      0.0882        &              &   (1)                \\
36114.6240   &    p   &   	pg     &    14465      &      0.0739        &              &   (1)                \\
36163.5630   &    p   &   	pg     &    14569      &      0.0619        &              &   (1)                \\
36607.4080   &    p   &   	pg     &    15512      &      0.0538        &              &   (1)                \\
37588.6110   &    s   &   	pg     &    17596.5    &      0.1202        &              &   (1)                \\
39531.3370   &    p   &   	pg     &    21724      &      0.1062        &              &   (1)                \\
39876.3610   &    p   &   	pg     &    22457      &      0.1203        &              &   (1)                \\
43112.7290   &    p   &   	pg     &    29333      &      0.0789        &              &   (1)                \\
43933.3740   &    s   &   	pg     &    31076.5    &      0.0898        &              &   (1)                \\
44171.5150   &    s   &   	pg     &    31582.5    &      0.0657        &              &   (1)               \\
44254.3890   &    s   &   	pg     &    31758.5    &      0.0997        &              &   (1)                \\
44256.4810   &    p   &   	pg     &    31763      &      0.0736        &              &   (1)               \\
44985.3470   &    s   &   	pg     &    33311.5    &      0.0886        &              &   (1)                \\
45676.5290   &    p   &   	pg     &    34780      &      0.0740        &              &   (1)                \\
46005.5320   &    p   &   	pg     &    35479      &      0.0703        &              &   (1)                \\
46036.3790   &    s   &   	pg     &    35544.5    &      0.0877        &              &   (1)                \\
46114.2760   &    p   &   	pg     &    35710      &      0.0868        &              &   (1)                \\
46321.6040   &    s   &   	pg     &    36150.5    &      0.0794        &              &   (1)               \\
46466.3390   &    p   &   	pg     &    36458      &      0.0796        &              &   (1)                \\
46707.6010   &    s   &   	pg     &    36970.5    &      0.1171        &              &   (1)                \\
46746.6460   &    s   &   	pg     &    37053.5    &      0.0955        &              &   (1)                \\
46768.5030   &    p   &   	pg     &    37100      &      0.0658        &              &   (1)                \\
50120.2312   &    p   &   	ccd    &    44221      &      0.0675        &   0.0025     &   (3)               \\
50138.3520   &    s   &   	ccd    &    44259.5    &      0.0670        &   0.0031     &   (3)              \\
50139.2915   &    s   &   	ccd    &    44261.5    &      0.0652        &   0.0020     &   (3)              \\
50147.2955   &    s   &   	ccd    &    44278.5    &      0.0676        &   0.0035     &   (3)              \\
50163.2988   &    s   &   	ccd    &    44312.5    &      0.0677        &   0.0021     &   (3)              \\
50773.5269   &    p   &   	ccd    &    45609      &      0.0566        &   0.0014     &   (3)             \\
50839.4256   &    p   &   	ccd    &    45749      &      0.0598        &   0.0021     &   (4)               \\
50863.4275   &    p   &   	ccd    &    45800      &      0.0569        &   0.0003     &   (5)              \\
50865.3103   &    p   &   	ccd    &    45804      &      0.0570        &   0.0005     &   (5)                \\
50888.3750   &    p   &   	ccd    &    45853      &      0.0583        &   0.0024     &   (4)                \\
50897.3183   &    p   &   	ccd    &    45872      &      0.0586        &   0.0003     &   (5)                \\
50904.3750   &    p   &   	ccd    &    45887      &      0.0551        &   0.0005     &   (6)               \\
51144.1871   &    s   &     ccd    &    46396.5    &      0.0547        &              &   (2)               \\
51144.8938   &    p   &     ccd    &    46398      &      0.0554        &              &   (2)               \\
51165.3671   &    s   &   	ccd    &    46441.5    &      0.0540        &   0.0002     &   (6)               \\
51165.6013   &    p   &   	ccd    &    46442      &      0.0529        &   0.0003     &   (6)             \\
51176.4278   &    p   &   	ccd    &    46465      &      0.0537        &   0.0003     &   (6)              \\
51225.3767   &    p   &   	ccd    &    46569      &      0.0516        &   0.0002     &   (6)              \\
51241.3810   &    p   &   	ccd    &    46603      &      0.0528        &   0.0020     &   (7)               \\
51481.4266   &    p   &   	ccd    &    47113      &      0.0505        &   0.0009     &   (7)             \\
51488.7227   &    s   &     ccd    &    47128.5    &      0.0511        &              &   (2)             \\
51543.5568   &    p   &   	ccd    &    47245      &      0.0507        &   0.0023     &   (7)             \\
51544.7311   &    s   &     ccd    &    47247.5    &      0.0483        &              &   (2)               \\
51568.2640   &    s   &   	ccd    &    47297.5    &      0.0471        &   0.0001     &   (8)                \\
51568.4986   &    p   &   	ccd    &    47298      &      0.0464        &   0.0007     &   (8)               \\
51571.3231   &    p   &   	ccd    &    47304      &      0.0468        &   0.0006     &   (8)                \\
51571.5555   &    s   &   	ccd    &    47304.5    &      0.0438        &   0.0035     &   (8)                \\
51586.6216   &    s   &     ccd    &    47336.5    &      0.0481        &              &   (2)                \\
51592.2700   &    s   &   	ccd    &    47348.5    &      0.0483        &   0.0023     &   (9)              \\
51602.6253   &    s   &     ccd    &    47370.5    &      0.0486        &              &   (2)              \\
51620.2764   &    p   &   	ccd    &    47408      &      0.0491        &   0.0024     &   (9)             \\
51626.3930   &    p   &   	ccd    &    47421      &      0.0469        &   0.0039     &   (10)               \\
51799.6057   &    p   &   	ccd    &    47789      &      0.0486        &   0.0022     &   (10)                \\
51881.7348   &    s   &     ccd    &    47963.5    &      0.0437        &              &   (2)                \\
51898.4460   &    p   &   	ccd    &    47999      &      0.0457        &   0.0016     &   (9)             \\
51912.5648   &    p   &     ccd    &    48029      &      0.0440        &              &   (2)               \\
51923.3900   &    p   &   	ccd    &    48052      &      0.0435        &   0.0021     &    (9)             \\
51924.3332   &    p   &   	ccd    &    48054      &      0.0454        &   0.0026     &    (11)               \\
51956.3386   &    p   &   	ccd    &    48122      &      0.0444        &   0.0025     &    (9)            \\
52209.5612   &    p   &   	ccd    &    48660      &      0.0401        &              &    (9)             \\
52279.4571   &    s   &   	ccd    &    48808.5    &      0.0397        &   0.0005     &    (8)            \\
52312.6396   &    p   &     ccd    &    48879      &      0.0391        &              &    (2)               \\
52321.3465   &    s   &   	ccd    &    48897.5    &      0.0384        &   0.0034     &    (9)            \\
52337.5850   &    p   &     ccd    &    48932      &      0.0384        &              &    (2)               \\
52609.8730   &    s   &     ccd    &    49510.5    &      0.0368        &              &    (2)               \\
52610.8135   &    s   &     ccd    &    49512.5    &      0.0360        &              &    (2)              \\
52619.5214   &    p   &   	ccd    &    49531      &      0.0363        &   0.0004     &    (12)            \\
52625.8747   &    s   &     ccd    &    49544.5    &      0.0354        &              &    (2)               \\
52662.5902   &    s   &     ccd    &    49622.5    &      0.0377        &              &    (2)               \\
52669.6486   &    s   &     ccd    &    49637.5    &      0.0358        &              &    (2)               \\
52672.4735   &    s   &   	ccd    &    49643.5    &      0.0366        &   0.0014     &    (11)            \\
52683.2987   &    s   &   	ccd    &    49666.5    &      0.0362        &   0.0039     &    (11)           \\
52683.3004   &    s   &   	ccd    &    49666.5    &      0.0379        &   0.0063     &    (11)            \\
52694.3591   &    p   &   	ccd    &    49690      &      0.0355        &   0.0042     &    (11)            \\
52694.3596   &    p   &   	ccd    &    49690      &      0.0360        &   0.0042     &    (11)            \\
52695.3015   &    p   &   	ccd    &    49692      &      0.0366        &   0.0030     &    (11)               \\
52695.3022   &    p   &   	ccd    &    49692      &      0.0373        &   0.0044     &    (11)               \\
52701.6544   &    s   &     ccd    &    49705.5    &      0.0353        &              &    (2)            \\
52723.3060   &    s   &   	ccd    &    49751.5    &      0.0355        &   0.0034     &    (11)            \\
52981.4722   &    p   &   	ccd    &    50300      &      0.0326        &   0.0001     &    (13)              \\
52983.3553   &    p   &   	ccd    &    50304      &      0.0330        &   0.0029     &    (11)           \\
52983.3557   &    p   &   	ccd    &    50304      &      0.0334        &   0.0029     &    (11)           \\
52983.3561   &    p   &   	ccd    &    50304      &      0.0338        &   0.0028     &    (11)            \\
52983.5916   &    s   &   	ccd    &    50304.5    &      0.0339        &   0.0029     &    (11)           \\
52983.5918   &    s   &   	ccd    &    50304.5    &      0.0341        &   0.0030     &    (11)           \\
52983.5933   &    s   &   	ccd    &    50304.5    &      0.0356        &   0.0029     &    (11)            \\
53035.6009   &    p   &   	ccd    &    50415      &      0.0329        &              &    (1)              \\
53047.1311   &    s   &   	ccd    &    50439.5    &      0.0314        &   0.0002     &    (14)               \\
53314.9481   &    s   &   	ccd    &    51008.5    &      0.0303        &              &    (1)           \\
53323.8904   &    s   &   	ccd    &    51027.5    &      0.0297        &              &    (1)               \\
53368.6039   &    s   &   	ccd    &    51122.5    &      0.0284        &              &    (1)               \\
53408.1405   &    s   &     ccd    &    51206.5    &      0.0277        &   0.0003     &    (16)              \\
53409.3196   &    p   &   	ccd    &    51209      &      0.0301        &   0.0004     &    (17)            \\
53411.2031   &    p   &     ccd    &    51213      &      0.0308        &   0.0003     &    (16)            \\
53413.7883   &    s   &   	ccd    &    51218.5    &      0.0273        &              &    (1)            \\
53435.6774   &    p   &   	ccd    &    51265      &      0.0297        &              &    (1)               \\
53445.3257   &    s   &   	ccd    &    51285.5    &      0.0290        &   0.0002     &    (17)           \\
53674.5457   &    s   &   	ccd    &    51772.5    &      0.0269        &   0.0005     &    (17)            \\
53717.3759   &    s   &   	ccd    &    51863.5    &      0.0250        &   0.0001     &    (18)               \\
53735.7347   &    s   &   	ccd    &    51902.5    &      0.0272        &              &    (1)              \\
53763.7384   &    p   &   	ccd    &    51962      &      0.0253        &              &    (1)               \\
54061.9119   &    s   &   	ccd    &    52595.5    &      0.0218        &              &    (1)               \\
54091.3295   &    p   &   	ccd    &    52658      &      0.0217        &   0.0016     &    (19)              \\
54091.5641   &    s   &   	ccd    &    52658.5    &      0.0210        &   0.0025     &    (19)            \\
54105.6856   &    s   &   	ccd    &    52688.5    &      0.0220        &              &    (1)            \\
54107.5678   &    s   &   	ccd    &    52692.5    &      0.0215        &              &    (1)            \\
54107.8033   &    p   &   	ccd    &    52693      &      0.0217        &              &    (1)             \\
54143.1051   &    p   &   	ccd    &    52768      &      0.0223        &              &    (1)            \\
54165.6978   &    p   &   	ccd    &    52816      &      0.0223        &              &    (1)             \\
54179.5823   &    s   &   	ccd    &    52845.5    &      0.0217        &              &    (1)             \\
54185.7023   &    s   &   	ccd    &    52858.5    &      0.0228        &              &    (1)             \\
54380.5621   &    s   &   	ccd    &    53272.5    &      0.0203        &   0.0002     &    (20)             \\
54476.3446   &    p   &   	ccd    &    53476      &      0.0190        &   0.0019     &    (21)            \\
54476.5794   &    s   &   	ccd    &    53476.5    &      0.0184        &   0.0010     &    (21)            \\
54496.5837   &    p   &   	ccd    &    53519      &      0.0187        &              &    (1)           \\
54500.3478   &    p   &   	ccd    &    53527      &      0.0174        &   0.0005     &    (21)           \\
54505.2898   &    s   &   	ccd    &    53537.5    &      0.0172        &   0.0005     &    (22)           \\
54520.5873   &    p   &   	ccd    &    53570      &      0.0176        &              &    (1)              \\
54526.7072   &    p   &   	ccd    &    53583      &      0.0186        &              &    (1)               \\
54800.8748   &    s   &   	ccd    &    54165.5    &      0.0139        &   0.0004     &    (23)               \\
54845.3560   &    p   &   	ccd    &    54260      &      0.0157        &   0.0005     &    (24)            \\
54846.0629   &    s   &     ccd    &    54261.5    &      0.0166        &              &    (2)            \\
54877.5967   &    s   &     ccd    &    54328.5    &      0.0147        &              &    (2)               \\
54888.6566   &    p   &     ccd    &    54352      &      0.0135        &   0.0004     &    (25)               \\
54890.7736   &    s   &     ccd    &    54356.5    &      0.0125        &   0.0004     &    (25)               \\
54905.6039   &    p   &   	ccd    &    54388      &      0.0163        &              &    (1)              \\
55114.8208   &    s   &     ccd    &    54832.5    &      0.0150        &              &    (2)               \\
55156.7104   &    s   &     ccd    &    54921.5    &      0.0139        &              &    (2)               \\
55181.4210   &    p   &   	ccd    &    54974      &      0.0137        &              &    (1)               \\
55206.6022   &    s   &     ccd    &    55027.5    &      0.0135        &              &    (2)               \\
55210.6022   &    p   &     ccd    &    55036      &      0.0127        &              &    (2)               \\
55253.6686   &    s   &     ccd    &    55127.5    &      0.0116        &              &    (2)               \\
55262.6120   &    s   &   	ccd    &    55146.5    &      0.0121        &              &   (1)                \\
55263.5537   &    s   &   	ccd    &    55148.5    &      0.0124        &              &   (1)                \\
55566.6698   &    s   &   	ccd    &    55792.5    &      0.0093        &              &   (1)                \\
55575.6091   &    s   &   	ccd    &    55811.5    &      0.0057        &   0.0005     &   (26)                \\
55577.9661   &    s   &     ccd    &    55816.5    &      0.0093        &   0.0003     &   (25)               \\
55581.9680   &    p   &     ccd    &    55825      &      0.0104        &   0.0004     &   (25)                \\
55631.6244   &    s   &   	ccd    &    55930.5    &      0.0098        &              &   (1)             \\
55896.8518   &    p   &   	ccd    &    56494      &      0.0079        &   0.0004     &   (27)                \\
55920.1498   &    s   &     ccd    &    56543.5    &      0.0071        &   0.0003     &   (25)             \\
55921.7978   &    p   &     ccd    &    56547      &      0.0078        &              &   (2)                \\
55958.9822   &    p   &     ccd    &    56626      &      0.0083        &   0.0002     &   (25)                 \\
55964.6295   &    p   &     ccd    &    56638      &      0.0074        &              &   (2)                \\
55981.3356   &    s   &     ccd    &    56673.5    &      0.0043        &              &   (2)                \\
56217.8539   &    p   &     ccd    &    57176      &      0.0049        &              &   (2)                \\
56221.8535   &    s   &     ccd    &    57184.5    &      0.0037        &              &   (2)                \\
56256.9186   &    p   &   	ccd    &    57259      &      0.0030        &   0.0003     &   (28)                \\
56256.9207   &    p   &     ccd    &    57259      &      0.0051        &              &   (2)               \\
56272.9166   &    p   &     ccd    &    57293      &      -0.0022       &   0.0017     &   (25)            \\
56272.9218   &    p   &   	ccd    &    57293      &      0.0030        &              &   (1)                \\
56309.6355   &    p   &   	ccd    &    57371      &      0.0034        &              &   (1)                \\
56325.6381   &    p   &   	ccd    &    57405      &      0.0029        &              &   (1)                \\
56592.9812   &    p   &     ccd    &    57973      &      -0.0014       &              &   (25)            \\
56592.9843   &    p   &   	ccd    &    57973      &      0.0017        &   0.0002     &   (1)                \\
56715.5916   &    s   &     ccd    &    58233.5    &      -0.0037       &   0.0001     &   (25)                 \\
56715.5940   &    s   &   	ccd    &    58233.5    &      -0.0013       &              &   (1)             \\
56956.8175   &    p   &   	ccd    &    58746      &      -0.0023       &              &   (1)                \\
57316.8890   &    p   &   	ccd    &    59511      &      -0.0025       &              &   (1)                \\
57320.8881   &    s   &   	ccd    &    59519.5    &      -0.0042       &              &   (1)                \\
57375.7239   &    p   &   	ccd    &    59636      &      -0.0029       &              &   (1)                \\
\hline\hline
\end{longtable}
\end{center}
\textbf
{\footnotesize Reference:} \footnotesize (1) The O-C gateway;
(2) BAV\footnote{http://http://www.bav-astro.eu/}; (3) \citet{2000IBVS.4887....1S}; (4) \citet{2000IBVS.4888....1S}; (5) \citet{1999IBVS.4712....1A};
(6) \citet{2001IBVS.5017....1A}; (7) \citet{2002IBVS.5263....1S}; (8) \citet{2002IBVS.5296....1A}; (9) \citet{2007OEJV...74....1B};
(10) \citet{2002IBVS.5287....1Z}; (11) \citet{2004IBVS.5583....1Z}; (12) \citet{2003IBVS.5484....1A}; (13) \citet{2006IBVS.5676....1K};
(14) \citet{2005IBVS.5592....1K}; (15) \citet{2005OEJV....3....1L}; (16) The present work; (17) \citet{2006IBVS.5741....1Z}; (18) \citet{2006IBVS.5731....1H};
(19) \citet{2007IBVS.5761....1H}; (20) \citet{2008IBVS.5835....1B}; (21) \citet{2009IBVS.5874....1H}; (22) \citet{2008IBVS.5837....1D};
(23) \citet{2009IBVS.5871....1D}; (24) \citet{2010IBVS.5918....1H}; (25) \citet{2017PASJ...69...69Y}; (26) \citet{2011IBVS.5992....1D};
(27) \citet{2012IBVS.6011....1D}; (28) \citet{2013IBVS.6042....1D};
\end{footnotesize}

\section{Analysis of Light Curves}

To model the light curves of GU Ori and determine its physical properties, the Wilson-Devinney (W-D) program \citep{Van2007,Wilson2012} is used. Since GU Ori has EW type light curves, Mode 3 for contact binary whose both component stars are filling their Roche lobe is applied. According to the spectral information obtained by LAMOST, the mean surface temperature of the star eclipsed at the primary minimum is set to be $T_1 = 6050K$. The gravity-darkening coefficients, the bolometric albedo coefficients and the limb darkening coefficients are adopted accordingly \citep{1967ZA.....65...89L,1969AcA....19..245R,1993AJ....106.2096V}.

\begin{figure}[!h]
\begin{center}
\includegraphics[width=12cm]{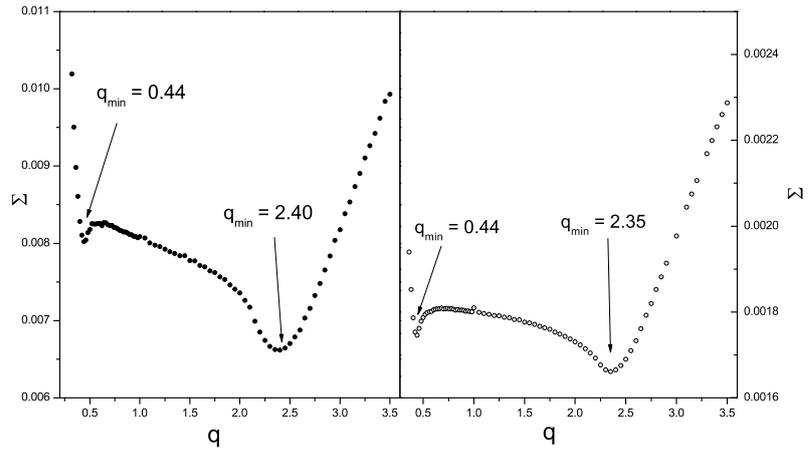}
\end{center}
\caption{The q-search diagram. Left panel: q - search results based on the data of the 1m telescope at Yunnan Observations, Chinese Academy of Sciences. Right panel: q - search results based on the data published by \citet{2017PASJ...69...69Y}.}\label{q-search}
\end{figure}

\begin{figure}[!ht]
\begin{center}
\includegraphics[width=10cm]{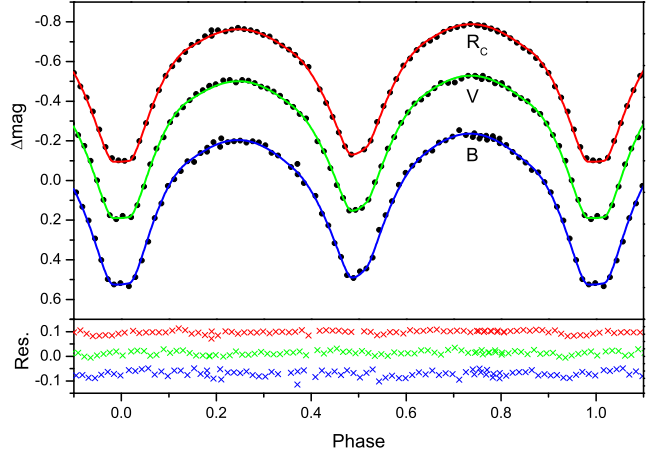}
\end{center}
\caption{The solid circles are the $B$ $V$ and $R_c$ bands light curves obtained by the 1m telescope at Yunnan Observations, Chinese Academy of Sciences. The color lines are the theoretical light curves. The color crosses in the lower part are the residuals.}\label{LC-zhou}
\end{figure}

First of all, the q-search method is used to determine the initial mass ratio of GU Ori, in which q is set from 0.02 to 9. The step is 0.02 while q ranges from 0.02 to 1 and 0.05 when q ranges from 1 to 9. The results are plotted in the left part of Fig. \ref{q-search}. It shows two minima at q = 0.44 and q = 2.40. Then the initial mass ratio of q = 2.40 is adopted and set as a free parameter. The final values of all adjustable parameters are listed in the second column of Table \ref{WD_results}. The corresponding theoretical light curves based on the determined parameters and the fitting results are displayed in Fig. \ref{LC-zhou}.

\begin{table}[!h]
\begin{center}
\caption{Photometric solutions of GU Ori}\label{WD_results}
\small
\begin{tabular}{lllllllll}
\hline\hline
Parameters                            &    1m data                       &    \citet{2017PASJ...69...69Y}        \\\hline
$T_{1}(K)   $                         &  6050(fixed)                     &  6050(fixed)      \\
q ($M_2/M_1$ )                        &  2.32($\pm0.05$)                 &  2.35($\pm0.06$)         \\
$i(^{\circ})$                         &  83.3($\pm0.4$)                  &  83.7($\pm0.5$)      \\
$\Omega_{in}$                         &  5.68                            &  5.68      \\
$\Omega_{out}$                        &  5.08                            &  5.08     \\
$\Omega_{1}=\Omega_{2}$               &  5.59($\pm0.07$)                 &  5.58($\pm0.09$)    \\
$T_{2}(K)$                            &  6021($\pm68$)                   &  5846($\pm16$)       \\
$\Delta T(K)$                         &  29                              &  204       \\
$T_{2}/T_{1}$                         &  0.995($\pm0.011$)               &  0.966($\pm0.003$)     \\
$L_{1}/(L_{1}+L_{2}$) ($B$)           &  0.326($\pm0.009$)               &                      \\
$L_{1}/(L_{1}+L_{2}$) ($V$)           &  0.325($\pm0.012$)               &  0.354($\pm0.001$)     \\
$L_{1}/(L_{1}+L_{2}$) ($R_c$)         &  0.324($\pm0.006$)               &  0.349($\pm0.001$)     \\
$r_{1}(pole)$                         &  0.297($\pm0.003$)               &  0.301($\pm0.003$)     \\
$r_{1}(side)$                         &  0.311($\pm0.004$)               &  0.315($\pm0.003$)     \\
$r_{1}(back)$                         &  0.350($\pm0.006$)               &  0.358($\pm0.004$)     \\
$r_{2}(pole)$                         &  0.434($\pm0.007$)               &  0.439($\pm0.009$)     \\
$r_{2}(side)$                         &  0.465($\pm0.009$)               &  0.471($\pm0.012$)     \\
$r_{2}(back)$                         &  0.495($\pm0.013$)               &  0.502($\pm0.017$)     \\
$f$                                   &  $17.7\,\%$($\pm$11.2\,\%$$)     &  $26.5\,\%$($\pm$14.8\,\%$$) \\
Spot 1                                &                                  &                              \\
$\theta(^{\circ})$                    &  142.2($\pm3.3$)                 &  7.3($\pm3.4$)   \\
$\psi(^{\circ})$                      &  33.6($\pm9.0$)                  &  275.4($\pm8.8$)  \\
$r$(rad)                              &  0.67($\pm0.05$)                 &  0.67($\pm0.15$)   \\
$T_f$                                 &  0.82(fixed)                     &  0.82(fixed)  \\
Spot 2                                &                                  &                           \\
$\theta(^{\circ})$                    &  36.8($\pm5.9$)                  &                         \\
$\psi(^{\circ})$                      &  188.1($\pm4.0$)                 &                        \\
$r$(rad)                              &  0.54($\pm0.14$)                 &                         \\
$T_f$                                 &  0.81(fixed    )                 &                         \\
$\Sigma{\omega(O-C)^2}$               &  0.000967                        &  0.001486            \\
\hline
\hline
\end{tabular}
\end{center}
\end{table}

\begin{figure}[!h]
\begin{center}
\includegraphics[width=10cm]{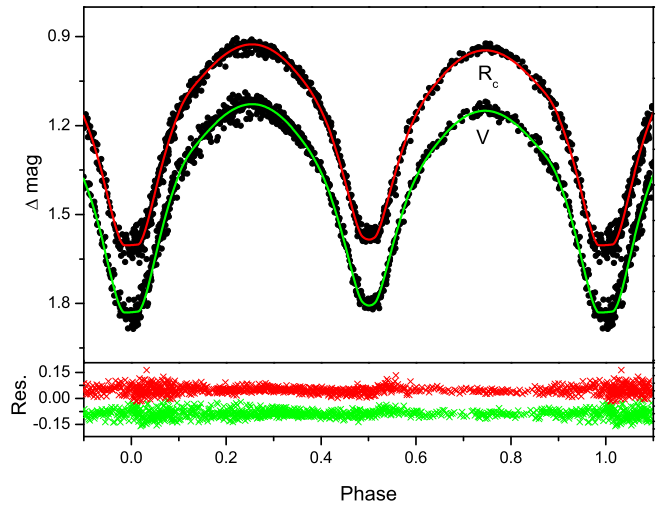}
\end{center}
\caption{The solid circles are the $V$ and $R_c$ bands light curves published by \citet{2017PASJ...69...69Y}. The green and red lines are the theoretical light curves. The color crosses in the lower part are the residuals.}\label{LC-yang}
\end{figure}

\citet{2017PASJ...69...69Y} also published the light curves' solution of GU Ori. They got the minimum at q = 0.45 when the q - search method was applied. Using the data published by them, we get the q - search diagram which are displayed in the right panel of Fig. \ref{q-search}. Two minima at q = 0.44 and q = 2.35 are obtained. It is consistent with the result determined using the light curves' data observed by the 1m telescope at Yunnan Observations, Chinese Academy of Sciences. The sulution is listed in the third column of Table \ref{WD_results}. The fitting light curves are shown in Fig. \ref{LC-yang}. We usually get two minima while the q - search method is applied and the two q values are inverse. The deeper minimum is usually chosen. The two ambiguity solutions correspond to A or W configuration and give out the same mass, radius and luminosity for the two components. However, the temperature and star eclipsed at the primary minimum are different. For q = 0.44, the massive star is mistaken as the star eclipsed at the primary minimum and the temperature of massive star is higher. In fact, it is a W-subtype system that the less massive but hotter star is the star eclipsed at the primary minimum.

\section{Discussions and Conclusions}

In the present work, two sets of light curves of GU Ori are analyzed with the Wilson-Devinney program, and its basic physical parameters are determined. GU Ori is a totally eclipsing binary, and the determined parameters have very high confidence \citep{2005Ap&SS.296..221T}. The mass ratio and orbital inclination calculated from the two sets of light curves are almost the same, which means the two separated observations give out consistent results. The fill-out factor determined by us is a little higher than that determined by \citet{2017PASJ...69...69Y}, which may be caused by mass transfer from star 2 to star 1. In 2005, the temperature difference between the two star is only 29$K$. However, the difference becomes quite larger (204$K$) after several years later. Since GU Ori is a solar-type contact binary, spots are common on its component stars. The spots' parameters are listed in Table \ref{WD_results}. The stellar activities may result in the changes of the determined mean surface temperature of the secondary star ($T_2$). The average temperature of star 2 is 5934(42)$K$, and the mass of star 2 is estimated to be 1.05 $M_\odot$ \citep{Cox2000}. Then, the absolute parameters of the two component stars are calculated and listed in Table \ref{elements}. The orbital semi-major axis is 2.92$R_\odot$.

\begin{table}[!h]
\caption{Absolute elements of  primary and secondary stars in GU Ori}\label{elements}
\begin{center}
\small
\begin{tabular}{ccc}
\hline
Parameters                        &    star 1                      &    star 2          \\
\hline
$M$                               & $0.45 M_\odot$                 & $1.05M_\odot$         \\
$R$                               & $0.95 R_\odot$                 & $1.37R_\odot$         \\
$L$                               & $1.09 L_\odot$                 & $2.10L_\odot$         \\
\hline
\end{tabular}
\end{center}
\end{table}

The O'Connell effect appears on late-type contact binaries commonly. It is suggested that it is closely related to magnetic activity of the component stars \citep{2016NewA...47....3Z,2015ApJ...805...22W}, circumstellar material around the binary system \citep{2003ChJAA...3..142L}, and mass transfer between the component stars \citep{1990BAAS...22.1296S}. For GU Ori, the negative O'Connell effect is reported on the light curves obtained in 2005, in which the light maxima after the secondary minima (Max II) is brighter than the light maxima at phase 0.25 (Max I). The difference between the two light maxima (Max I - Max II) are 0.037 mag in $B$ band, 0.033 mag in $V$ band, 0.030 mag in $R_c$ band. However, the O'Connell effect changed to positive one and the Max I - Max II are -0.026 mag in $V$ band, -0.022 mag in $R_c$ band for the light curves obtained by \citet{2017PASJ...69...69Y}. The O'Connell effect on GU Ori may be related to stellar activitives\citep{2016PASJ...68..102X}.

The period investigation of \citet{2017PASJ...69...69Y} shown that the period of GU Ori was increasing at a rate of $dP/dt=+1.45\times{10^{-7}}day\cdot year^{-1}$. However, only times of light minimum observed by CCD method were used in their work. In our O - C diagram, times of light minima obtained by photograph method are also used. Although the pg data don't have as high accuracy as the CCD data, they do show a downward tendency. While the variation tendency is upward when only CCD data are used. Therefore, the pg data are very important while the period variation of GU Ori is investigated since they will give a quite different result, as show in Fig. \ref{O-C}. The mass transfer from the more massive star to the smaller one is supposed to explain the long term period decreasing ($dP/dt=-6.24\times{10^{-8}}day\cdot year^{-1}$), with its rate to be $\frac{dM_{2}}{dt}= - 2.98\times{10^{-8}}M_\odot/year$. And also, a cyclic period change is revealed, which may be caused by the light-travel time effect of a tertiary component.

The W UMa type contact binary stars may evolve from short period detached binaries, and it will take a quite long pre-contact evolution time scale \citep{2006AcA....56..199S,2011AcA....61..139S,2013MNRAS.430.2029Y,2014MNRAS.437..185Y}. The nuclear evolution and angular momentum evolution have played very important role during the evolution stage from detached to contact status \citep{1988MNRAS.231..341H}. Thus, most of research work have been focusing on the two evolutionary mechanisms. Nowadays, the LAMOST have obtained atmosphere parameters for thousands of contact binaries, which attract our research interest. The statistical research shows that the atmosphere parameters of contact binaries may also give us some important information on their formation and evolution. The high metallicity is a key factor to select a group of binary systems, in which GU Ori is a sample. \citet{2018ApJS..235....5Q} pointed out that contact binary with high metallicity may be contaminated by compact object during its evolution lifetime. In the case of UX Eri ([Fe/H] = 0.45), the mass of the tertiary is estimated to be less than $0.56 M_\odot$ \citep{2007AJ....134.1769Q}. The tertiary component may have shortened its pre-contact evolution time \citep{2013ApJS..209...13Q}. GU Ori is also a contact binary with quite high metallicity ([Fe/H] = 0.31). Its formation and evolution may be different from that of UX Eri. The cyclic period change implies that there may be a tertiary component orbiting GU Ori. However, third light isn't detected through the light curves. The information is insufficient to estimate the parameters of the tertiary component by far, and more evidences are needed to confirm whether GU Ori is really contaminated by the tertiary component. The present work is just the beginning, and more and more contact binaries with high metallicity will be investigated carefully in the future, which will shed new light on research of contact binaries.

\bigskip

\vskip 0.3in \noindent
We thank the referee for valuable comments that have improved the quality of the present work. This work is partly supported by the Chinese Natural Science Foundation (Grant No. 11703080 and 11703082) and the Yunnan Natural Science Foundation (No. 2018FB006). Guoshoujing Telescope (the Large Sky Area Multi-Object Fiber Spectroscopic Telescope LAMOST) is a National Major Scientific Project built by the Chinese Academy of Sciences. Funding for the project has been provided by the National Development and Reform Commission. LAMOST is operated and managed by the National Astronomical Observatories, Chinese Academy of Sciences. This work is part of the research activities at the National Astronomical Research Institute of Thailand (Public Organization).


\end{document}